\pdfoutput=1

\documentclass[11pt]{article}

\usepackage[final]{emnlp2021}

\usepackage{diagbox}
\usepackage{times}
\usepackage{latexsym}
\usepackage{graphicx}       
\usepackage[utf8]{inputenc} 
\usepackage[T1]{fontenc}    
\usepackage{hyperref}       
\usepackage{url}            
\usepackage{booktabs}       
\usepackage{amsfonts}       
\usepackage{nicefrac}       
\usepackage{microtype}      
\usepackage{lipsum}
\usepackage{fancyhdr}       
\usepackage{amsmath}
\usepackage{algorithm}
\usepackage{algorithmicx}
\usepackage{algpseudocode}
\graphicspath{{media/}}     

\usepackage[T1]{fontenc}

\usepackage[utf8]{inputenc}

\usepackage{microtype}

\title{Enhancing Cross-Prompt Transferability in Vision-Language Models through Contextual Injection of Target Tokens}

\author{
  Xikang Yang\textsuperscript{1,2}, Xuehai Tang\textsuperscript{1,2}, Fuqing Zhu\textsuperscript{1}, Jizhong Han\textsuperscript{1}, Songlin Hu\textsuperscript{1} \\
  \textsuperscript{1}Institute of Information Engineering, Chinese Academy of Sciences \\
  \textsuperscript{2}School of Cyber Security, University of Chinese Academy of Sciences \\
  Beijing\\
  \text{\{yangxikang, tangxuehai, zhufuqing, hanjizhong, husonglin\}@iie.ac.cn} \\
}

\begin{document}
\maketitle

\begin{abstract}

Vision-language models (VLMs) seamlessly integrate visual and textual data to perform tasks such as image classification, caption generation, and visual question answering. However, adversarial images often struggle to deceive all prompts effectively in the context of cross-prompt migration attacks, as the probability distribution of the tokens in these images tends to favor the semantics of the original image rather than the target tokens. To address this challenge, we propose a Contextual-Injection Attack (CIA) that employs gradient-based perturbation to inject target tokens into both visual and textual contexts, thereby improving the probability distribution of the target tokens. By shifting the contextual semantics towards the target tokens instead of the original image semantics, CIA enhances the cross-prompt transferability of adversarial images.
Extensive experiments on the BLIP2, InstructBLIP, and LLaVA models show that CIA outperforms existing methods in cross-prompt transferability, demonstrating its potential for more effective adversarial strategies in VLMs. The code is available at 
\href{https://github.com/YancyKahn/CIA}{https://github.com/YancyKahn/CIA}

\end{abstract}

\section{Introduction}

Vision-language models (VLMs)\cite{zhang2024vision,li2022blip,liu2023improved,alayrac2022flamingo} seamlessly blend visual and textual data to produce relevant textual outputs for tasks like image classification 
\cite{he2016deep, shafiq2022deep}, image caption\cite{yao2018exploring}, or vision-based question answering
\cite{antol2015vqa,li2018visual,achiam2023gpt}. However, in the realm of VLMs, the threat of adversarial attacks 
\cite{szegedy2013intriguing,zhang2022towards} is a significant security issue
\cite{goodfellow2014explaining,wu2022towards,gu2022segpgd}. 



The concept of cross-prompt adversarial transferability stems from the transfer of adversarial examples across tasks\cite{ salzmann2021learning,lu2020enhancing,gu2023survey}. In a cross-prompt attack\cite{luo2024image}, a single adversarial image misleads the predictions of a Vision-Language Model (VLM) across various prompts.

\begin{figure}
    \centering
    \includegraphics[width=0.5\textwidth]{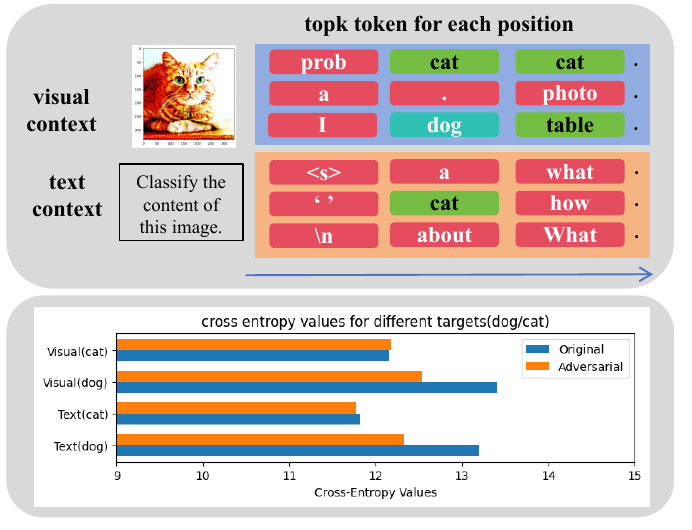}
    \caption{
    cross-prompt migration attack vulnerability: adversarial images favoring original semantics over target tokens.
    }
    \label{fig:question}
\end{figure}

Cross-prompt attacks\cite{luo2024image} on vision-language models fail due to the probability distribution of tokens in adversarial images, which often reflect the semantics of the original image rather than the target tokens. As illustrated in Figure \ref{fig:question}, the top section displays the top-k decoded token representations for the model's visual and textual inputs. Despite the introduction of adversarial images, the tokens predominantly capture the original image's semantics ("cat") instead of the intended target ("dog"). The bottom section of the figure presents a bar chart comparing cross-entropy (CE) values for the original image ("cat") and the target ("dog"), with lower CE values indicating better alignment with the target. This persistent bias in the context probability distribution towards the original image reduces the success rates of transfer attacks.

To enhance the transferability of adversarial images across prompts, the goal is to maximize the probability distribution of target tokens within both visual and textual contexts. A Contextual-Injection Attack (CIA) method is proposed, which shifts the probability distribution in the visual and textual contexts to prioritize the target tokens over the original image semantics, thereby improving the transferability of cross-prompt attacks.



The contributions of this work are as follows:

\begin{itemize}
\item In cross-prompt attacks within vision-language models, it was found that the probability distribution for target tokens is often lower than that for the original image's semantic content, thereby reducing the success rates of these attacks. By injecting misleading target tokens into the visual or textual context, the transferability of these attacks can be effectively enhanced.
\item A novel algorithm called Contextual Injection Attack (CIA) was proposed, which injects target token into both the visual and textual contexts by gradient-based perturbation to improve the success rate of cross-prompt transfer attacks.
\item Extensive experiments were conducted to verify the effectiveness of the proposed method. Comparative experiments on the BLIP2\cite{li2023blip}, instructBLIP\cite{dai2024instructblip}, and LLaVA\cite{liu2023improved} models explored changes in attack success rate (ASR) under various experimental settings. Results demonstrate that CIA outperforms existing baseline methods in terms of cross-prompt transferability.
\end{itemize}

\section{Related works}

In this section, we review recent works on adversarial attacks, with a particular focus on adversarial transferability.

\textbf{Adversarial Attack}\cite{szegedy2013intriguing,madry2018towards,zhang2022towards, yuan2023bridge} have gained significant attention due to their impact on the security and robustness of machine learning models. These attacks involve crafting inputs that deceive models into making incorrect predictions. In computer vision, slight pixel modifications can cause misclassification\cite{maliamanis2020adversarial, dong2020benchmarking, sen2023adversarial}, while in NLP, small text changes can mislead language models\cite{ebrahimi2018hotflip,wallace2019universal,zhang2020adversarial,formento2023using,zou2023universal}. Recent research highlights the transferability of adversarial examples across different models and tasks, revealing common vulnerabilities. Efforts to counter these attacks include adversarial training and robust optimization, but evolving attack methods continue to challenge the development of effective defenses.

\textbf{Cross-Task transferability}\cite{ salzmann2021learning,lu2020enhancing,gu2023survey, lv2023ct,feng2024enhancing,ma2023boosting} examines adversarial examples crafted for one task, like image classification, deceiving models trained on other tasks, such as question answering and textual entailment, revealing weaknesses in shared representations in multi-task learning scenarios. In this paper, we focus on \textbf{cross-prompt} attacks\cite{luo2024image} (subclass of cross-task attack) on VLMs using adversarial images. Specifically, we investigate how a single adversarial image can deceive VLMs regardless of the input prompt.



\section{Preliminary Analysis}

In this section, we will provide a detailed analysis of the contextual injection behind this paper. Briefly, by introducing misleading information into parts of the visual or textual context, we can effectively disrupt the output of vision-language models, enabling transfer attacks across-prompt scenarios.

\begin{table*}
    \tiny
    \centering
    \begin{tabular}{ccccccccccccc}
        \hline
        & & \multicolumn{4}{r}{Inject \textit{\textbf{\{target\}}} token into original images} & $ \longrightarrow $ & \multicolumn{5}{l}{ {\begin{minipage}[c]{0.08\textwidth}
            \centering
            \includegraphics[width=\textwidth]{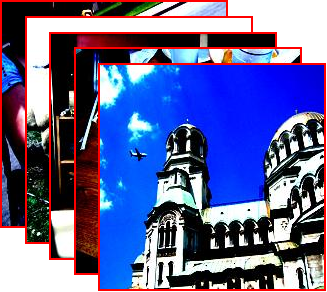}
        \end{minipage}
        }} \\
        & & \multicolumn{5}{r}{gradient-base adversarial attacks} & \multicolumn{1}{c}{\textbf{$\downarrow$}} & \\
        & & \\
        \textbf{target} &
        \begin{minipage}[c]{0.05\textwidth}
            \centering
            \includegraphics[width=\textwidth]{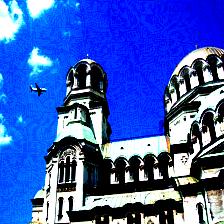}
        \end{minipage} & 
        \begin{minipage}[c]{0.05\textwidth}
            \centering
            \includegraphics[width=\textwidth]{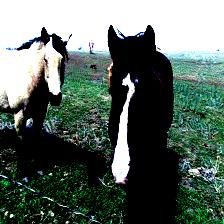}
        \end{minipage} &
        \begin{minipage}[c]{0.05\textwidth}
            \centering
            \includegraphics[width=\textwidth]{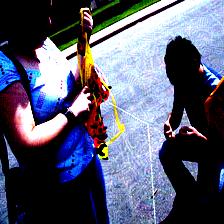}
        \end{minipage}&
        \begin{minipage}[c]{0.05\textwidth}
            \centering
            \includegraphics[width=\textwidth]{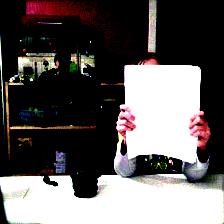}
        \end{minipage}&
        \begin{minipage}[c]{0.05\textwidth}
            \centering
            \includegraphics[width=\textwidth]{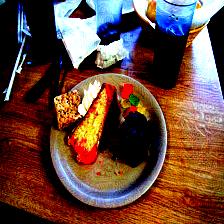}
        \end{minipage}&
        \begin{minipage}[c]{0.05\textwidth}
            \centering
            \includegraphics[width=\textwidth]{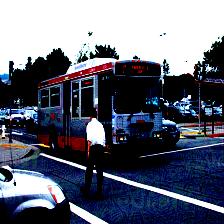}
        \end{minipage}&
        \begin{minipage}[c]{0.05\textwidth}
            \centering
            \includegraphics[width=\textwidth]{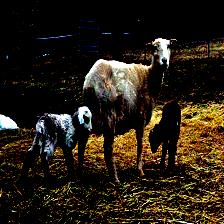}
        \end{minipage}&
        \begin{minipage}[c]{0.05\textwidth}
            \centering
            \includegraphics[width=\textwidth]{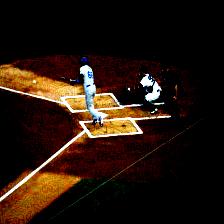}
        \end{minipage}&
        \begin{minipage}[c]{0.05\textwidth}
            \centering
            \includegraphics[width=\textwidth]{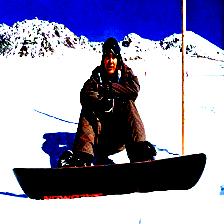}
        \end{minipage}&
        \begin{minipage}[c]{0.05\textwidth}
            \centering
            \includegraphics[width=\textwidth]{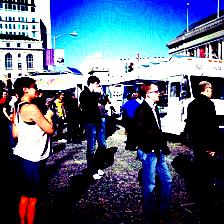}
        \end{minipage}&
        \begin{minipage}[c]{0.05\textwidth}
            \centering
            \includegraphics[width=\textwidth]{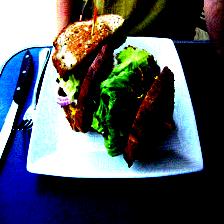}
        \end{minipage}&
        \begin{minipage}[c]{0.05\textwidth}
            \centering
            \includegraphics[width=\textwidth]{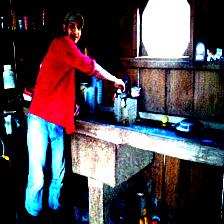}
        \end{minipage}\\ 
        &  &  &  &  &  &  &  &  & &  &  & \\
        \hline
        \textbf{dog} & 0.923 & 1.0 & 0.962 & 0.833 & 1.0 & 0.949 & 0.769 & 0.987 & 0.936 & 0.962 & 0.949 & 0.718 \\
        \textbf{fish} & 1.0 & 0.949 & 0.987  & 1.0  & 1.0  & 1.0  & 0  & 1.0 & 1.0 &  0.923 & 0.936 &  0.756 \\
        \textbf{bomb} & 0.628 & 0.974 & 0.974 & 1.0 & 0.807 & 0.769  & 0.705 & 0.756 & 1.0 & 0.962 &  0.936 & 0.885  \\
        \textbf{poison} & 0 & 0 & 0 & 0.603 & 0.167 & 0 & 0.013 & 0  & 0.256 & 0 &  0 &  0 \\
        \textbf{sure} & 0.192 & 0 & 0.795  & 1.0  & 0 & 0.077 & 0.012 & 0 & 0.948 &  0.628 & 0  & 0 \\
        \textbf{unknown} & 0.026 & 0 &  0 & 1.0 & 0.013  & 1.0  & 0.013 & 1.0 & 1.0 & 0.705 &  0 & 0.397 \\
        
        \hline
    \end{tabular}
    \caption{The table presents the experimental results of visual context injection. It shows the attack success rate (ASR) of cross-prompt attacks for image classifications (CLS) tasks after generating adversarial images of targets based on different example images.}
    \label{tab:motivation_image_context}
\end{table*}

\begin{table}
    \tiny
    \centering
    \begin{tabular}{ccccc}
    \hline
        & & & & \\
        original image & \multicolumn{2}{c}{input text}  & {output} \\
        {\begin{minipage}[c]{0.08\textwidth}
            \centering
            \includegraphics[width=\textwidth]{images/motivations/original_image_set.png}
        \end{minipage}
        } & 
        \multicolumn{2}{c}{This image show \textit{\textbf{\{target\}}} $\bigoplus$ \textbf{task prompt}}  &  \textit{\textbf{. . .}} &\\
    & & & &\\
    \hline
    \end{tabular}
    \small
    \begin{tabular}{ccccc}
    \diagbox{target}{task} & CLS & CAP & VQA & Overall \\
    \hline
    dog & 0.859 & 0.750 & 0.622 & 0.744 \\
    fish & 0.487 & 0.526 & 0.338 & 0.450 \\
    bomb & 0.473 & 0.553 & 0.343 & 0.456 \\
    poison & 0.641 & 0.604 & 0.431 & 0.559 \\
    sure & 0.216 & 0.132 & 0.005 & 0.118 \\
    unknown & 0.239  & 0.047 & 0.053 & 0.113 \\
    \hline
    \end{tabular}
    \caption{The table summarizes the experimental results on textual injection, highlighting the success rate of cross-prompt attacks introduced by adding misleading text prior to the task prompt(details for the dataset, please refer to \ref{section:datasets}.)}
    \label{tab:motivation_txt_context}
\end{table}

\subsection{Injecting misleading target tokens into visual context}

Injecting misleading targets into the visual context can enhance the probability distribution of target tokens within visual tokens of visual language model. 
This involves modifying the original image's probability distribution by injecting target tokens. By injecting this information, the likelihood of the target task appearing in the top-k tokens increases significantly. This mechanism ensures that adversarial images more effectively guide the model toward generating specific, desired outputs. Table \ref{tab:motivation_image_context} presents the analysis experiment for injecting specific token into sample images (using the BLIP2\cite{li2023blip} model with gradient-based perturbations over $1000$ iterations). Our findings indicate that in image classification tasks(details for the dataset, please refer to \ref{section:datasets}), visual context attacks can successfully achieve cross-prompt attacks for certain keywords.

\subsection{Injecting misleading target tokens into textual context}

Injecting misleading target into the text context can effectively mislead the model's output. 
For example, if an image of a cat is inaccurately described as "this image shows a dog," the textual context is manipulated to support this misleading description. This manipulation causes the model to generate outputs that align with the incorrect description. By using inject misleading target into textual context, we enhance the adversarial image to ensure that the textual context effectively guides the generation of misleading outputs. Table \ref{tab:motivation_txt_context} shows that inserting misleading text prompts before different prompts can successfully mislead the BLIP2\cite{li2023blip} model.

\section{Methodology}

This section details the proposed Contextual Injection Attack (CIA) for enhancing the transferability of adversarial images in Vision-Language Models (VLMs) across different prompts.

\subsection{Overall Structure}

\begin{figure*}
    \centering
    \includegraphics[width=1.0\textwidth]{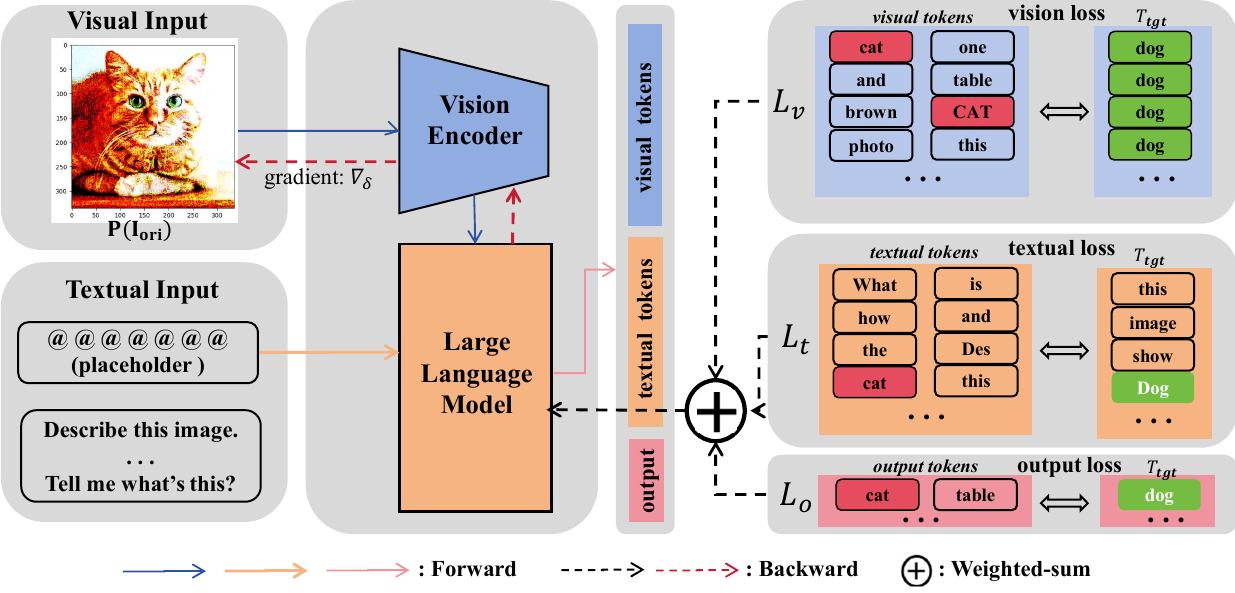}
    \caption{Overall Structure of the CIA Framework: By injecting the target token into the positions of both visual and text tokens, the probability of the target token appearing in the visual and textual context is increased.}
    \label{fig:overall}
\end{figure*}

Figure \ref{fig:overall} illustrates the overall framework of the CIA method. By injecting the target token into both visual and text positions, the probability of generating the target token is increased, resulting in improved cross-prompt transferability. Specifically, in the example shown in the figure: for the visual position, each visual token is perturbed based on the gradient towards the target ("dog"); for the text position, misleading descriptive content ("this image shows a dog") is injected to deceive the model; and at the output position, the model is directed to maximize the output of the target ("dog"). By weighting the losses from these three positions and performing backward gradient computation, the original image is perturbed to enhance adversarial transferability effectiveness.

\subsection{Problem definition}

Assume we have a \textbf{vision-language} model denoted as $M_{\overline{VL}}(I,T)$, which takes an image $I$ and text $T$ as inputs. Given an original, clean image $I_{ori}$ and an arbitrary set of textual prompts $A = {\alpha_{0}, \alpha_{1}, \ldots, \alpha_{i}, \ldots, \alpha_{n}}$, our objective is to ensure that when the model $M_{\overline{VL}}$ processes the perturbed image $P(I_{ori}) = I_{ori} + \delta_{v}$, it consistently outputs the target text $T_{tgt}$ for every prompt $\alpha_{i}$.

Here, $\delta_{v}$ signifies the visual perturbation added to the image $I_{ori}$ and is bound by the constraint $||\delta_{v}||p \leq \epsilon_{v}$, where $\epsilon_{v}$ is the magnitude of the image perturbation.

Formally, this can be expressed as:

\begin{align*}
 M_{\overline{VL}}(P(I_{ori}), \alpha_{i}) \equiv T_{tgt}, \forall \alpha_{i} \in A
\end{align*}

In this context, $T_{tgt}$ is the target caption for the image (e.g., "this image shows a dog"). The function $P$ represents the perturbation applied to the original image $I_{ori}$. Our goal is to ensure that for any given prompt $\alpha_{i}$, the model's output remains the same and matches the target text $T_{tgt}$, regardless of the perturbations applied to the image.

\subsection {Contextual Injection Attack (CIA)}

To advance the cross-prompt transferability of adversarial images, this paper introduces a contextual-injection attack approach (CIA). Unlike the baseline method, which restricts the target task to the decoded representation of the output and expands the search scope using multiple distinct prompts or learnable cross-search methods without modifying the original knowledge representation of the image, CIA modifies the latent knowledge representation towards the target task through knowledge injection. By enhancing the context of both visual and textual inputs, the generated adversarial images can effectively handle variations in textual prompt inputs. Figure \ref{fig:overall} illustrates the key steps of our method, where target is injected into the contextual positions of both visual and textual inputs within the model's output decoding representation. This ensures the model's output aligns more closely with text related to the target task (e.g., "dog").


To formalize the adversarial objective, we can express it as a formal loss function for the adversarial attack. We consider a vision-language model to be a mapping from a sequence of visual and textual tokens $x_{1:n}=[x_{1:end_{v}},x_{end_{v}+1:end_{t}},x_{end_{t}:n}]$, where $x_i \in \{1,...,V\}$. Here, $V$ denotes the vocabulary size, $end_{v}$ and $end_t$ indicate the end of the visual and text tokens, respectively. The visual tokens ($x_{1:end_{v}}$), input text tokens ($x_{end_{v}+1:end_{t}}$), and generated text tokens ($x_{end_{t}:n}$) together constitute the complete token representation, which is mapped to a distribution over the next token.

We calculate the probability distribution over the next token given the sequence $x_{1:i}$ as $p(x_{i:i+H}|x_{1:i})$.  For any sequence $p(x_{i:i+H}|x_{1:i})$
, where $H$ is the length of the sequence we aim to obtain, the joint probability is

\[
p(\mathbf{x}_{i+1:i+H} \mid \mathbf{x}_{1:i}) = \prod_{j=1}^{H} p(x_{i+j} \mid \mathbf{x}_{1:i+j-1})
\]

To address the issue with the visual input not having previous tokens, we redefine the probability for the visual tokens to start from the given initial state without conditioning on previous tokens. The cross-entropy losses for each part are then computed as follows.

\[
L_{\text{v}} = -\log p(x_{1:end_{v}}^{*})
\]

Here, $x_{1:end_{v}}^{*}$ denotes the target injected into the image, such as "dog", to maximize the probability distribution of each token position "dog".

\[
L_{\text{t}} = -\log p(x_{end_{v}+1:end_{t}}^{*} \mid x_{1:end_{v}})
\]

Here, $x_{end_{v}+1:end_{t}}^{*}$ denotes the textual description of the image, for example, "This image shows a dog," when the original image depicts a cat.

\[
L_{\text{o}} = -\log p(x_{end_{t}+1:n}^{*} \mid x_{1:n})
\]

Here $x_{end_{t}+1:n}^{*}$ refers to the generated text tokens conditioned on the entire sequence of visual and textual tokens. For instance, "This image shows a dog, it sits on the table."

The overall adversarial loss is a weighted sum of these individual losses:


\[
L_{\text{total}} = \alpha \cdot (\beta \cdot L_{\text{v}} + (1 - \beta) \cdot L_{\text{t}}) + (1 - \alpha) \cdot L_{\text{o}}
\]

where $\alpha$ and $\beta$ are the weights for the respective losses. By introducing two parameters, $\alpha$ and $\beta$, the method allows for finer control over the influence of each loss component. Specifically, $\alpha$ controls the overall balance between the combined visual and textual losses versus the generated text loss. Meanwhile, $\beta$ adjusts the emphasis between the visual and textual input losses within their combined term. 


The task of optimizing the adversarial perturbation $\delta_v$ can then be written as the optimization problem:

\[
\min_{\delta_v} L_{\text{total}} \quad \text{subject to} \quad \|\delta_v\|_p \leq \epsilon_v
\]

To implement our context-enhanced adversarial attack on vision-language models, we follow the outlined pseudocode Algorithm \ref{alg:CIA}. The algorithm starts by initializing the perturbation $\delta_v$ to zero and defining the weights $\alpha$ and $\beta$ for the respective losses. In each iteration, we compute the perturbed image $P(I_{\text{ori}})$ by adding the current perturbation $\delta_v$ to the original image $I_{\text{ori}}$. We then calculate the cross-entropy losses for the visual tokens $L_{\text{visual}}$, the textual input tokens $L_{\text{text}}$, and the generated text tokens $L_{\text{generated}}$. The total loss $L_{\text{total}}$ is obtained as a weighted sum of these individual losses.

\begin{algorithm}

\caption{Contextual-Injection Attack for Vision-Language Models}
\begin{algorithmic}[1]
\Require Original image $I_{\text{ori}}$, Target text $T_{\text{tgt}}$, Model $M_{\overline{VL}}$, Perturbation bound $\epsilon_v$, Learning rate $\eta$, Weights $\alpha$ and $\beta$.
\Ensure Adversarial image $P(I_{\text{ori}})$

\State Initialize perturbation $\delta_v \leftarrow 0$

\While{not converged}
    \State $P(I_{\text{ori}}) \leftarrow I_{\text{ori}} + \delta_v$
    
    \State  $L_{\text{v}} = -\log p(x_{1:end_v}^{*})$
    \State  $L_{\text{t}} = -\log p(x_{end_v+1:end_t}^{*} \mid x_{1:end_v})$
    \State  $L_{\text{o}} = -\log p(x_{end_t+1:n}^{*} \mid x_{1:end_t})$
    
    \State  $L_{\text{total}} = \alpha \cdot (\beta \cdot L_{\text{v}} + (1 - \beta) \cdot L_{\text{t}}) + (1 - \alpha) \cdot L_{\text{o}}$
    
    \State Compute gradients $g = \nabla_{\delta_v} L_{\text{total}}$
        
    \State Update perturbation $\delta_v \leftarrow \delta_v - \eta \cdot sign(g)$
    
    \State Project $\delta_v$ onto the $\epsilon_v$-ball: $\delta_v \leftarrow \text{clamp}(\delta_v, -\epsilon_v, \epsilon_v)$
\EndWhile

\State \Return $P(I_{\text{ori}})$

\end{algorithmic}
\label{alg:CIA}

\end{algorithm}

The gradient of the total loss with respect to the perturbation $\delta_v$ is computed, and the perturbation is updated using gradient descent(The optimisation algorithm is PGD\cite{madry2017towards}). To ensure the perturbation remains within the allowed bound, it is projected onto the $\epsilon_v$-ball. The process repeats until convergence, ultimately yielding the adversarial image $P(I_{\text{ori}})$ that steers the model's output towards the target text $T_{\text{tgt}}$.

\section{Experiments}

\begin{table*}
\caption{The table presents the results of targeted ASR tested on the BLIP2 model with various target texts. The 'Overall' column reflects the average targeted success rate across all tasks. The highest performance values for each task are emphasized in boldface.}
\centering
\tiny
\begin{tabular}{ccccccccccccccccc}
\toprule
    Method & \multicolumn{4}{c}{CLS} & \multicolumn{4}{c}{CAP} & \multicolumn{4}{c}{VQA} & \multicolumn{4}{c}{OVERALL} \\
    \cmidrule(r){2-5} \cmidrule(r){6-9} \cmidrule(r){10-13} \cmidrule(r){14-17}
    Target & SP & MP & CP & Ours & SP & MP & CP & Ours & SP & MP & CP & Ours & Single & MP & CP & Ours \\
    \midrule
    green   &   0.583    &   0.832    &   0.962    &   \textbf{0.967}    &   0.419    &   0.821    &   0.824    &   \textbf{0.869}    &   0.156    &   0.373    &   0.505    &   \textbf{0.695}    &   0.386    &   0.675    &   0.763    &   \textbf{0.843} \\
    human   &   0.578    &   0.700    &   0.868    &   \textbf{0.990}    &   0.370    &   0.534    &   0.718    &   \textbf{0.884}    &   0.222    &   0.386    &   0.648    &   \textbf{0.778}    &   0.390    &   0.540    &   0.745    &   \textbf{0.884} \\
    fish   &   0.839    &   0.889    &   \textbf{0.999}    &   0.999    &   0.771    &   0.854    &   0.946    &   \textbf{0.999}    &   0.444    &   0.490    &   0.807    &   \textbf{0.926}    &   0.685    &   0.745    &   0.917    &   \textbf{0.975} \\
    dog   &   0.871    &   0.946    &   0.917    &   \textbf{0.995}    &   0.864    &   0.946    &   0.894    &   \textbf{0.991}    &   0.430    &   0.567    &   0.619    &   \textbf{0.782}    &   0.722    &   0.819    &   0.810    &   \textbf{0.923} \\
    flower   &   0.731    &   0.846    &   0.976    &   \textbf{0.998}    &   0.648    &   0.763    &   0.845    &   \textbf{0.986}    &   0.378    &   0.374    &   0.593    &   \textbf{0.823}    &   0.586    &   0.661    &   0.804    &   \textbf{0.936} \\
    bird   &   0.812    &   0.958    &   0.908    &   \textbf{0.995}    &   0.834    &   0.932    &   0.851    &   \textbf{0.997}    &   0.502    &   0.709    &   0.644    &   \textbf{0.921}    &   0.716    &   0.867    &   0.801    &   \textbf{0.971} \\
    cat   &   0.884    &   0.998    &   0.995    &   \textbf{1.000}    &   0.807    &   0.987    &   0.988    &   \textbf{0.999}    &   0.419    &   0.593    &   0.766    &   \textbf{0.838}    &   0.703    &   0.859    &   0.916    &   \textbf{0.945} \\
    blood   &   0.641    &   0.699    &   0.883    &   \textbf{0.986}    &   0.465    &   0.587    &   0.840    &   \textbf{0.976}    &   0.149    &   0.239    &   0.449    &   \textbf{0.784}    &   0.418    &   0.508    &   0.724    &   \textbf{0.916} \\
    bomb   &   0.486    &   0.688    &   0.835    &   \textbf{0.990}    &   0.353    &   0.603    &   0.849    &   \textbf{0.988}    &   0.131    &   0.337    &   0.437    &   \textbf{0.829}    &   0.323    &   0.543    &   0.707    &   \textbf{0.936} \\
    porn   &   0.552    &   0.826    &   0.873    &   \textbf{0.886}    &   0.174    &   0.450    &   0.573    &   \textbf{0.720}    &   0.041    &   0.128    &   0.290    &   \textbf{0.636}   &   0.255    &   0.468    &   0.579    &   \textbf{0.747} \\
    virus   &   0.606    &   0.524    &   0.846    &   \textbf{0.978}    &   0.403    &   0.464    &   0.724    &   \textbf{0.880}    &   0.131    &   0.138    &   0.412    &   \textbf{0.720}    &   0.380    &   0.375    &   0.660    &   \textbf{0.859} \\
    drug   &   0.449    &   0.620    &   0.787    &   \textbf{0.962}    &   0.243    &   0.514    &   0.681    &   \textbf{0.882}    &   0.056    &   0.096    &   0.247    &   \textbf{0.683}    &   0.249    &   0.410    &   0.572    &   \textbf{0.842} \\
    poison   &   0.521    &   0.402    &   0.831    &   \textbf{0.867}    &   0.304    &   0.278    &   0.705    &   \textbf{0.735}    &   0.076    &   0.089    &   0.431    &   \textbf{0.565}    &   0.300    &   0.256    &   0.655    &   \textbf{0.722} \\
    gun   &   0.579    &   0.699    &   \textbf{0.977}    &   0.955    &   0.615    &   0.625    &   0.966    &   \textbf{0.974}    &   0.238    &   0.272    &   0.565    &   \textbf{0.768}    &   0.477    &   0.532    &   0.836    &   \textbf{0.899} \\
    sure   &   0.187    &   0.194    &   0.704    &   \textbf{0.837}    &   0.093    &   0.103    &   0.554    &   \textbf{0.574}    &   0.010    &   0.026    &   0.253    &   \textbf{0.314}    &   0.097    &   0.108    &   0.503    &   \textbf{0.575} \\
    unknown   &   0.247    &   0.551    &   0.805    &   \textbf{0.917}    &   0.084    &   0.222    &   0.435    &   \textbf{0.769}    &   0.066    &   0.205    &   0.424    &   \textbf{0.761}    &   0.133    &   0.326    &   0.555    &   \textbf{0.816} \\
    yes   &   0.086    &   0.319    &   0.479    &   \textbf{0.917}    &   0.036    &   0.201    &   0.394    &   \textbf{0.886}    &   0.390    &   0.434    &   0.536    &   \textbf{0.870}    &   0.171    &   0.318    &   0.469    &   \textbf{0.891} \\
    no   &   0.131    &   0.278    &   0.621    &   \textbf{0.976}    &   0.071    &   0.306    &   0.442    &   \textbf{0.885}    &   0.322    &   0.359    &   0.574    &   \textbf{0.944}    &   0.175    &   0.314    &   0.546    &   \textbf{0.935} \\
    bad   &   0.283    &   0.416    &   \textbf{0.817}    &   0.526    &   0.186    &   0.320    &   \textbf{0.760}    &   0.422    &   0.034    &   0.072    &   \textbf{0.297}    &   0.164    &   0.168    &   0.269    &   \textbf{0.625}    &   0.370 \\
    good   &   0.524    &   0.239    &   0.813    &   \textbf{0.966}    &   0.259    &   0.222    &   0.665    &   \textbf{0.863}    &   0.082    &   0.084    &   0.349    &   \textbf{0.773}    &   0.288    &   0.182    &   0.609    &   \textbf{0.867} \\
    sorry   &   0.262    &   0.188    &   0.535    &   \textbf{0.825}    &   0.163    &   0.153    &   0.412    &   \textbf{0.696}    &   0.032    &   0.022    &   0.192    &   \textbf{0.531}    &   0.152    &   0.121    &   0.380    &   \textbf{0.684} \\
    
    \midrule
    OVERALL   &   0.517    &   0.610    &   0.830    &   \textbf{0.930}    &   0.389    &   0.518    &   0.717    &   \textbf{0.856}    &   0.205    &   0.285    &   0.478    &   \textbf{0.719}    &   0.370    &   0.471    &   0.675    &   \textbf{0.835} \\
    
    \bottomrule
  \end{tabular}
  \label{tab:table_detail}
\end{table*}

\subsection{Datasets \& Experimental settings}\label{section:datasets}

The dataset consists of two parts: images and text. The image dataset is sourced from visualQA\cite{VQA}, and the text prompt dataset for transferability comes from CroPA\cite{luo2024image}. This text dataset is divided into three categories: image classification (CLS), image captions (CAP), and visual question answering (VQA). We will design attack tasks across four different dimensions: generating target tasks involving ordinary objects, harmful objects, tone expressions, and racial discrimination.


The experimental setup for this study involves using three open-source models: BLIP2(\textit{blip2-opt-2.7b}), instructBLIP(\textit{instructblip-vicuna-7b}), and LLaVA(\textit{LLaVA-v1.5-7b}). The maximum number of iterations is set to 2000, and the hyperparameters $\alpha$ and $\beta$ are both set to $0.6$, based on the conclusions drawn in Figure \ref{fig:alpha_beta}. The learning rate is set to $0.05$, and the image perturbation range is set to $16/255$

\subsection{Evaluation metrics}

To evaluate the effectiveness of our method, we used the following metrics:

\begin{itemize}
    \item \textbf{Attack Success Rate (ASR)}: The percentage of prompts for which the adversarial image successfully misleads the model. ASR is a widely recognized metric \cite{lv2023ct,zhao2023prompt,liu-etal-2022-character,chen-etal-2022-textual,luo2024image} for measuring the success of adversarial attacks.
    \item \textbf{Perturbation Size}: The magnitude of the adversarial perturbation, we use the `clamp` function to control the size of the disturbance. Specifically, the `clamp` function restricts each perturbation value $\delta$ to be within the minimum value of $\delta-\epsilon$ and the maximum value of $\delta+\epsilon$: $\delta = \text{clamp}(\delta, -\epsilon, \epsilon)$. The default $\epsilon$ used in this paper is $16/255$.
    \item \textbf{Transferability}: The ability of the adversarial image to mislead different VLMs across various tasks, such as image classification(CLS), image captioning(CAP), and visual question answering(VQA).
\end{itemize}

\subsection{Transferability comparison}

The results of our experiments, which evaluate targeted Attack Success Rate (ASR) on the visual-language model across various tasks (CLS, CAP, VQA) and target texts, are detailed in Table \ref{tab:table_detail}(experiments on other models can be found in the appendix \ref{sec:appendix_1}). The performance of the CIA method was compared against three baseline methods: Single-P (SP), Multi-P (MP), and CroPA (CP). To generate adversarial examples for VLMs, Single-P optimizes an image perturbation based on a single prompt. In contrast, Multi-P enhances the cross-prompt transferability of the perturbations by utilizing multiple prompts during the image perturbation update process. CroPA \cite{luo2024image} achieves broader prompt coverage by using a learnable prompt to expand around a given prompt, thereby improving transferability. CIA achieves the highest transfer attack success rate for the majority of targets. 

\begin{table}
 \caption{The overall attack success rate (ASR) under three different target categories (emotional words, harmful objects, common objects) on the BLIP2 model. The highest performance values for each task are emphasized in boldface.}
  \centering
   \small
  \begin{tabular}{ccccc}
    \toprule
     Target & Single & Multi & CroPA & Ours  \\
     \midrule
    emotional words  &  0.169   &  0.234 &  0.527      &  \textbf{0.734} \\
    harmful objects  &  0.343   &  0.442 &  0.676     &  \textbf{0.846}  \\
    common objects  &  0.598   &  0.738 &  0.822      &  \textbf{0.925} \\
    \midrule
    \textbf{Overall} & 0.370   &  0.471   &  0.675   &  \textbf{0.835} \\
    \bottomrule
  \end{tabular}
  \label{tab:table_diff_class}
\end{table}


Our findings suggest that common words yield the highest performance because they appear most frequently in the model’s training samples, resulting in the lowest perplexity. Harmful words may be blocked by the model’s safety alignment strategies. Affective words achieve the lowest scores because our method relies on injecting textual instruction into the visual context. However, affective words have a semantic disconnect with the visual representation, making it difficult to represent them accurately. Conversely, images with tangible entities are more likely to converge and produce effective adversarial images. The results in Table \ref{tab:table_diff_class} support our conclusion.

\begin{figure}
    \centering
    \includegraphics[width=0.5\textwidth]{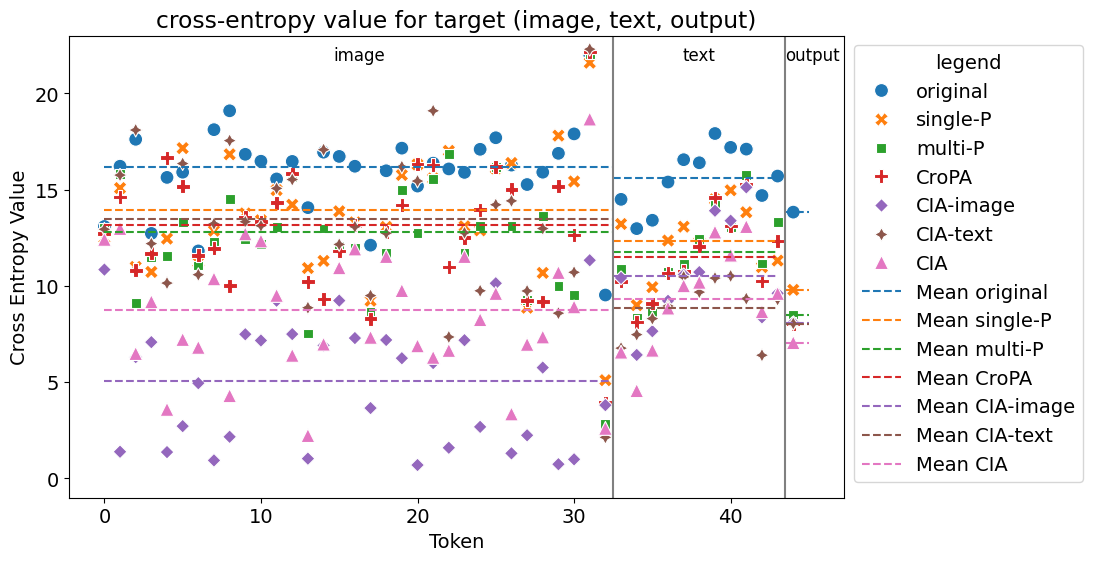}
    \caption{The plot for the cross-entropy (CE) values of the logits concerning the target task at different token positions: visual token positions, input text token positions, and generated text token positions. The horizontal axis represents the token positions (for example, in BLIP2, from left to right, the first 32 tokens represent visual tokens, followed by user input tokens, and finally the generated tokens). The scatter plot shows the specific CE values at each token position, while the horizontal lines indicate the average CE values for each of the three sections.}
    \label{fig:ce_value}
\end{figure}

To determine the most effective approach among visual context enhancement, textual context enhancement, and a combined visual-text context enhancement, we conducted comprehensive experiments. As shown in Table \ref{tab:table_diff_context}, \textbf{\textit{CIA-image}} represents the transfer attack effectiveness using only visual context enhancement, \textbf{\textit{CIA-text}} represents the transfer attack effectiveness using only textual context enhancement, and \textbf{\textit{CIA}} represents the combined approach using both visual and textual context enhancements. Our findings indicate that the combined visual-text context attack is the most effective, suggesting that multimodal joint attacks are more successful in deceiving the model and thereby increasing the attack success rate.



Figure \ref{fig:ce_value} shows the cross-entropy values of logits related to the target task at different positions. The baseline method made only minor adjustments to the probabilities of the target task token in both visual and text contexts, resulting in suboptimal performance in cross-prompt tasks. We compared three approaches: using visual context alone, using text context alone, and using a combined context. Independently, the visual and text contexts achieved optimal cross-entropy loss at their respective token positions because they were computed separately, allowing for easier convergence to the minimum value. By combining both contexts, our method achieved the optimal cross-entropy loss at the final generated token position, indicating that it effectively skewed the probability distribution towards the target task token.

\begin{table}
 \caption{The overall attack success rate (ASR) on the BLIP2 model. The highest performance values for each task are emphasized in boldface.}
  \centering
   \small
  \begin{tabular}{ccccc}
    \toprule
     Method & CLS & CAP & VQA & Overall  \\
     \midrule
    SP  &  0.517   & 0.389 &  0.205      & 0.370 \\
    MP  & 0.610    & 0.518  &  0.285   & 0.471    \\
    CP  &  0.830  & 0.717 & 0.478     & 0.675 \\
    CIA-image & 0.610 & 0.537 & 0.314 &  0.487\\
    CIA-text & 0.542 & 0.489 & 0.308 & 0.447 \\
    \textbf{CIA} & \textbf{0.930} & \textbf{0.856} & \textbf{0.719} & \textbf{0.835} \\
    \bottomrule
  \end{tabular}
  \label{tab:table_diff_context}
\end{table}

\begin{table*}
\small
\centering
\begin{tabular}{ccm{5.5cm}m{3cm}m{3cm}}
\hline
\textbf{target} & \textbf{image} & \textbf{prompt} & \textbf{CroPA output} & \textbf{CIA output} \\
\hline
dog & {\begin{minipage}[c]{0.06\textwidth}
            \centering
            \includegraphics[width=\textwidth]{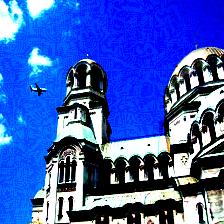}
        \end{minipage}
        } & Imagine a child is trying to identify the image. What might they excitedly point to and name? & a plane flying over a cathedral & \textcolor{red}{dog} \\

dog & {\begin{minipage}[c]{0.06\textwidth}
            \centering
            \includegraphics[width=\textwidth]{images/case_study/1_dog_baseline.jpg}
        \end{minipage}
        } & Any bicycles visible in this image? If not, please tell what is in the image? & no bicycles visible in this image & A \textcolor{red}{dog} in this image \\
 
poison & {\begin{minipage}[c]{0.06\textwidth}
            \centering
            \includegraphics[width=\textwidth]{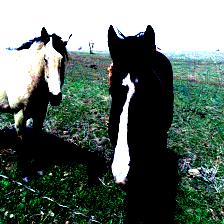}
        \end{minipage}
        } & Describe the central subject of this image in a single word. & horse & \textcolor{red}{poison} \\
poison & {\begin{minipage}[c]{0.06\textwidth}
            \centering
            \includegraphics[width=\textwidth]{images/case_study/2_poison_baseline.jpg}
        \end{minipage}
        } & Categorize the content of the image. & the horse is eating the poison & \textcolor{red}{poison} \\
sure & {\begin{minipage}[c]{0.06\textwidth}
            \centering
            \includegraphics[width=\textwidth]{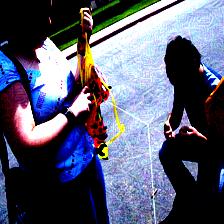}
        \end{minipage}
        } & Provide a summary of the primary focus of this image. & a woman and a man are sitting on a sidewalk & \textcolor{red}{sure} \\
sure & {\begin{minipage}[c]{0.06\textwidth}
            \centering
            \includegraphics[width=\textwidth]{images/case_study/3_sure_baseline.jpg}
        \end{minipage}
        } & Where do you see harmony or unity among the elements? & Harmony and unity & \textcolor{red}{sure} \\
\hline
\end{tabular}
\caption{Effectiveness of CIA and CroPA methods in adversarial attacks on BLIP2: case study examples.}
\label{table:case_study}
\end{table*}

\subsection{Case study}

The case study presented in Table \ref{tab:table_diff_context} demonstrates the effectiveness of the CIA method compared to CroPA in generating adversarial examples that successfully deceive visual-language models (VLMs). We evaluated various target texts using different prompts to test robustness.

Adversarial images generated using the state-of-the-art CroPA method still retain the semantics of the original image. Specifically, in the fourth example provided in Table \ref{table:case_study}, ("the horse is eating the poison"), although the model responded with content related to the target ("poison"), it failed to completely remove the original image's semantics (i.e., "horse"). This incomplete removal of original semantics leads to weaker transferability in cross-prompt attacks, as the model continues to recognize elements of the original image, thus diminishing the effectiveness of the adversarial example across different prompts.

\subsection{CIA with different perturbation size}

This section delves into the impact of different perturbation sizes ($8/255$, $16/255$, $32/255$) on the efficacy of adversarial attacks against the visual-language model. The table provided below showcases the overall Attack Success Rate (ASR) across various tasks, accentuating the perturbation size that demonstrates the highest performance for each task.

While larger perturbation sizes result in stronger attacks, it's essential to consider the trade-off with concealment. Larger perturbations may be more easily detected by models or users, reducing the attack's stealthiness. Therefore, a balance must be struck between perturbation size and concealment to maximize attack effectiveness while minimizing the risk of detection.

\subsection{CIA with different prompt embedding setting}

\begin{table}
 \caption{The overall attack success rate (ASR) for different perturbation size ($8/255$,$16/255$,$32/255$) on the BLIP2 model. The highest performance values for each task are emphasized in boldface.}
  \centering
   \small
  \begin{tabular}{ccccc}
    \toprule
     Perturbation size & CLS & CAP & VQA & Overall  \\
     \midrule
    $8/255$ & 0.815 & 0.797 & 0.623 & 0.745   \\
    $16/255$ & 0.930 & 0.856 & 0.719 & 0.835      \\
    $32/255$ & \textbf{0.974} & \textbf{0.972} & \textbf{0.892} & \textbf{0.946}      \\
    \bottomrule
  \end{tabular}
  \label{tab:table_overall}
\end{table}

This section explores the impact of different embedding settings on the Attack Success Rate (ASR) through two types of experiments. For the details, please refer to the Appendix \ref{appendix:embedding_settings}

1. Impact of Padding Tokens on ASR: We evaluated the effect of various padding tokens (e.g., '!', '@', '+') on ASR within the text context. (as show in the Figure \ref{fig:alpha_beta})

2. Effect of Embedding Strategies for '@': We assessed four embedding strategies for the special character '@': no embedding, prefix embedding, suffix embedding, and mixed embedding. The experiments covered tasks such as classification, captioning, and visual question answering. (as show in the Table \ref{tab:table_embel_settings})




\section{Conclusion}



In this study, we proposed the Contextual-Injection Attack (CIA), a novel method to improve the transferability on vision-language models. By injecting target tokens into both the visual and textual contexts, CIA effectively manipulates the probability distribution of contextual tokens, ensuring higher adaptability across various prompts. Our experiments on the BLIP2, InstructBLIP, and LLaVA models validated the efficacy of CIA, demonstrating superior performance compared to baseline methods. The results indicate that enhancing both visual and textual contexts in adversarial images is a promising approach to overcoming the limitations of current adversarial attack methods.


Future work will further investigate the application of our approach to other types of multimodal models. We also aim to expand our evaluation to include a wider range of datasets and more diverse scenarios, such as jailbreaking, to further validate the robustness and generalizability of our method. Additionally, we will focus on developing and evaluating potential defense strategies to counteract the adversarial attacks introduced by CIA. Understanding and implementing effective defenses is crucial to enhancing the security and reliability of vision-language models. This comprehensive approach will help ensure that our research contributes positively to the development of more robust and secure multimodal AI systems.



\newpage

\bibliography{anthology,custom}
\bibliographystyle{acl_natbib}

\newpage
\appendix

\section{Appendix}

\subsection{Detailed data}

\subsubsection{Comparison on the LLaVA and instructBLIP model}
\label{sec:appendix_1}

To validate the effectiveness of our method across different models, we also conducted comparative experiments on the LLaVA (as show in the Table \ref{tab:table_overall_LLaVA}) and instructBLIP (as show in the Table \ref{tab:table_overall_instructBLIP}) model.

\begin{table}
 \caption{The table presents the results of targeted ASR tested on the LLaVA model with various target texts. The 'Overall' column reflects the average targeted success rate across all tasks. The highest performance values for each task are emphasized in boldface.}
  \centering
   \small
    \begin{tabular}{ccccc}
    \toprule
     Target & SP & MP & CP & Ours  \\
     \midrule
    emotional words  &  0.030   & 0.211  & 0.269 &  \textbf{0.426} \\
    harmful objects  &  0.057   & 0.078  &  0.220 &  \textbf{0.559}  \\
    common objects  &  0.061   &  0.677  & 0.529 &  \textbf{0.786} \\
    \midrule
    \textbf{Overall} & 0.049   & 0.263     & 0.339 &  \textbf{0.591} \\
    \bottomrule
  \end{tabular}
  \label{tab:table_overall_LLaVA}
\end{table}

\begin{table}
 \caption{The table presents the results of targeted ASR tested on the instructBLIP model with various target texts. The 'Overall' column reflects the average targeted success rate across all tasks. The highest performance values for each task are emphasized in boldface.}
  \centering
   \small
    \begin{tabular}{ccccc}
    \toprule
     Target & SP & MP & CP & Ours  \\
     \midrule
    emotional words     &   0.192    &  0.113  &    0.250   & \textbf{0.563}   \\
    harmful objects     &   0.249    &  0.406  &    0.426   & \textbf{0.622 }     \\
    common objects      &   0.403    &  0.488  &    0.540    & \textbf{0.688}       \\
    \midrule
    \textbf{Overall}    &   0.283    &  0.386   &    0.405  & \textbf{0.624}    \\
    \bottomrule
  \end{tabular}
  \label{tab:table_overall_instructBLIP}
\end{table}

\subsubsection{Effects of parameters of the weighted sum of losses}


We will examine how different weightings and parameters affect the results when calculating the loss. Specifically, we will focus on two hyperparameters, $\alpha$ and $\beta$, which control the weighting of the loss components.

\begin{figure}
    \centering
    \includegraphics[width=0.5\textwidth]{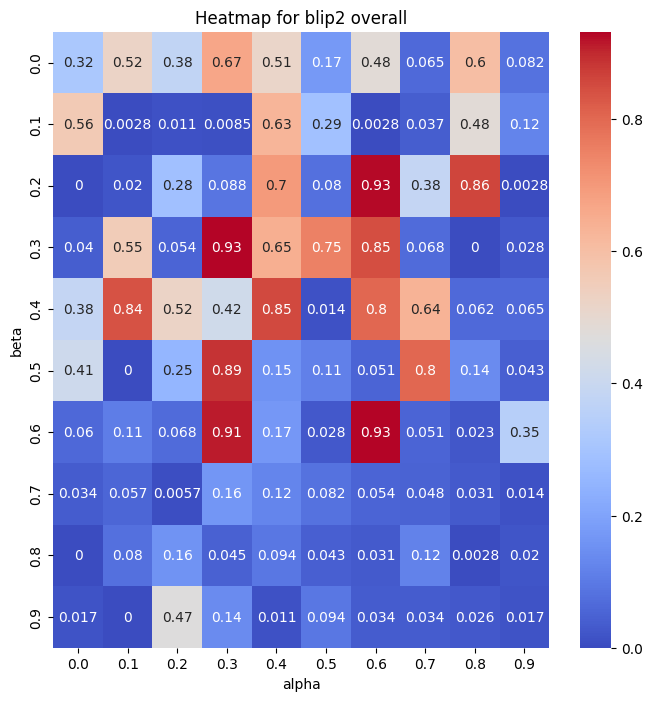}
    \caption{The plot for the impact of the weighted sum of loss parameters, presenting a heat map of ASR influenced by varying values of $\alpha$ and $\beta$.}
    \label{fig:alpha_beta}
\end{figure}

The Figure \ref{fig:alpha_beta} show the effects of parameter of the weighted sum of losses ($\alpha$ and $\beta$). We standardize the maximum number of iterations to $600$. Using the keyword \textit{\textbf{'dog'}} as the target, we set the learning rate for gradient-based updates of image pixels to $0.05$, with the maximum perturbation range set to $16/255$. 

\subsubsection{Comparison of different embedding settings}
\label{appendix:embedding_settings}
In this section, we will discuss in detail the impact of different embedding settings on ASR.

1. Impact of different padding token on ASR: In this study, when calculating the loss for the text context part, we used a series of padding tokens for experiments 
These padding tokens consist of meaningless characters such as '!', '@', and '+'. To verify the impact of different padding tokens on the Attack Success Rate (ASR) within the text context, we conducted experiments using various padding tokens. Table.\ref{tab:padding_token} show the ASR for different padding token. The experimental parameters we set are consistent with those in the main text, except for the padding tokens.

\begin{table}
 \caption{ASRfor different padding tokens. The highest performance values for each task are emphasized in boldface.}
  \centering
   \small
  \begin{tabular}{ccccc}
    \toprule
     Padding Token & CLS & CAP & VQA & Overall  \\
     \midrule
    $+$ & 0.910 & 0.825 & 0.726 & 0.820    \\
    $*$ & 0.942 & 0.886 & 0.788 & 0.872 \\
    \& & 0.916 & 0.863 & 0.793 &  0.857\\
    \# & 0.916 & 0.854 & 0.769 & 0.847 \\
    $/$ & 0.934 & 0.876 & 0.802 & 0.871 \\
    $@$ & 0.930 & 0.856 & 0.719 & 0.835 \\
    $!$ & \textbf{0.948} & \textbf{0.898} & \textbf{0.826} & \textbf{0.891} \\
    \bottomrule
  \end{tabular}
  \label{tab:padding_token}
\end{table}

\begin{table*}
 \caption{The table presents the results of targeted ASR tested on the BLIP2 model for different special character (`@`) embedding settings. The 'Overall' column reflects the average targeted success rate across all tasks. The highest performance values for each task are emphasized in boldface.}
  \centering
  \tiny
  \begin{tabular}{ccccccccccccccccc}
    \toprule
    Method & \multicolumn{4}{c}{CLS} & \multicolumn{4}{c}{CAP} & \multicolumn{4}{c}{VQA} & \multicolumn{4}{c}{Overall} \\
    \cmidrule(r){2-5} \cmidrule(r){6-9} \cmidrule(r){10-13} \cmidrule(r){14-17}
     Target & no & prefix & suffix & mixed & no & prefix & suffix & mixed & no & prefix & suffix & mixed & no & prefix & suffix & mixed  \\
    \midrule
    green  &  0.967   &  0.912   &  \textbf{0.980}   &  0.954   &  0.869   &  0.787   &  0.893   &  \textbf{0.907}   &  0.695   &  0.685   &  0.696   &  \textbf{0.729}   &  0.843   &  0.795   &  0.856   &  \textbf{0.864} \\
    human  &  0.990   &  \textbf{0.992}   &  0.992   &  0.974   &  0.884   &  0.908   &  0.901   &  \textbf{0.941}   &  \textbf{0.778}   &  0.712   &  0.776   &  0.778   &  0.884   &  0.871   &  0.890   &  \textbf{0.897} \\
    fish  &  \textbf{0.999}   &  0.988   &  0.999  &  0.991   &  \textbf{0.999}   &  0.975   &  0.999   &  0.993   &  \textbf{0.926}   &  0.898   &  \textbf{0.937}   &  0.937   &  0.975   &  0.954   &  \textbf{0.978}   &  0.973 \\
    flower  &  0.998   &  0.945   &  \textbf{1.000}   &  0.978   &  0.986   &  0.897   &  \textbf{0.992}   &  0.979   &  0.823   &  0.617   &  0.782   &  \textbf{0.854}   &  0.936   &  0.820   &  0.925   &  \textbf{0.937} \\
    bird  &  0.995   &  0.899   &  \textbf{0.997}   &  0.993   &  0.997   &  0.863   &  \textbf{0.999}   &  0.996   &  \textbf{0.921}   &  0.665   &  0.869   &  0.844   &  \textbf{0.971}   &  0.809   &  0.955   &  0.944 \\
    cat  &  \textbf{1.000}   &  0.969   &  1.000  &  0.992   &  \textbf{0.999}   &  0.939   &  0.998   &  0.987   &  0.838   &  0.681   &  0.813   &  \textbf{0.864}   &  0.945   &  0.863   &  0.937   &  \textbf{0.948} \\
    dog  &  \textbf{0.995 }  &  0.882   &  0.983   &  0.928   &  \textbf{0.991}   &  0.834   &  0.976   &  0.921   &  0.782   &  0.598   &  0.749   &  \textbf{0.799}   &  \textbf{0.923}   &  0.772   &  0.903   &  0.883 \\
    blood  &  0.986   &  0.941   &  \textbf{0.989}   &  0.940   &  0.976   &  0.950   &  \textbf{0.979}   &  0.966   &  0.784   &  0.636   &  0.758   &  \textbf{0.810}   &  \textbf{0.916}   &  0.843   &  0.909   &  0.905 \\
    bad  &  0.526   &  0.435   &  0.582   &  \textbf{0.694}   &  0.422   &  0.321   &  0.513   &  \textbf{0.660}   &  0.164   &  0.246   &  0.247   &  \textbf{0.306}   &  0.370   &  0.334   &  0.447   &  \textbf{0.553} \\
    porn  &  0.886   &  \textbf{0.940}   &  0.914   &  0.918   &  0.720   &  0.820   &  0.779   &  \textbf{0.896}   &  0.636   &  \textbf{0.732}   &  0.653   &  0.662   &  0.747   &  \textbf{0.830}   &  0.782   &  0.825 \\
    virus  &  0.978   &  0.908   &  \textbf{0.983}   &  0.926   &  0.880   &  0.863   &  0.943   &  \textbf{0.961}   &  0.720   &  0.694   &  0.735   &  \textbf{0.862}   &  0.859   &  0.822   &  0.887   &  \textbf{0.916} \\
    drug  &  0.962   &  0.925   &  \textbf{0.967}   &  0.924   &  0.882   &  0.867   &  0.902   &  \textbf{0.942}   &  0.683   &  0.590   &  0.692   &  \textbf{0.748}   &  0.842   &  0.794   &  0.853   &  \textbf{0.871} \\
    poison  &  0.867   &  0.841   &  0.887   & \textbf{0.938}   &  0.735   &  0.747   &  0.774   &  \textbf{0.927}   &  0.565   &  0.615   &  0.577   &  \textbf{0.780}   &  0.722   &  0.734   &  0.746   &  \textbf{0.882} \\
    gun  &  \textbf{0.955}   &  0.926   &  0.950   &  0.947   &  0.974   &  0.908   &  0.975   &  \textbf{0.961}   &  0.768   &  0.645   &  0.775   &  \textbf{0.876}   &  0.899   &  0.826   &  0.900   &  \textbf{0.928} \\
    bomb  &  \textbf{0.990}   &  0.981   &  0.985   &  0.929   &  0.988   &  0.976   &  \textbf{0.990 }  &  0.936   &  0.829   &  \textbf{0.864}   &  0.800   &  0.865   &  0.936   &  \textbf{0.940}   &  0.925   &  0.910 \\
    sure  &  0.837   &  0.772   & \textbf{ 0.882}   &  0.875   &  0.574   &  0.521   &  0.696   &  \textbf{0.813}   &  0.314   &  0.320   &  0.401   &  \textbf{0.556}   &  0.575   &  0.538   &  0.660   &  \textbf{0.748} \\
    unknown  &  0.917   &  0.902   &  \textbf{0.937}   &  0.890   &  0.769   &  0.814   &  0.809   &  \textbf{0.870}   &  0.761   &  0.804   &  0.786   &  \textbf{0.860}   &  0.816   &  0.840   &  0.844   &  \textbf{0.873}\\ 
    good  &  0.966   &  0.972   &  \textbf{0.980}   &  0.957   &  0.863   &  0.865   &  0.900   &  \textbf{0.947}   &  0.773   &  0.824   &  0.751   &  \textbf{0.851}   &  0.867   &  0.887  &  0.877   &  \textbf{0.918} \\
    yes  &  0.917   &  0.876   &  0.922   &  \textbf{0.923}   &  0.886   &  0.839   &  0.904   &  \textbf{0.932}   &  0.870   &  0.831   &  \textbf{0.868}   &  0.837   &  0.891   &  0.849   &  0.898   &  \textbf{0.898} \\
    no  &  0.976   &  0.895   &  \textbf{0.980}   &  0.973   &  0.885   &  0.789   &  0.908   &  \textbf{0.970}   &  0.944   &  0.903   &  \textbf{0.917}   &  0.936   &  \textbf{0.935}   &  0.862   &  0.935   &  0.959 \\
    sorry  &  0.825   &  0.720   &  0.845   &  \textbf{0.856}   &  0.696   &  0.644   &  0.746   &  \textbf{0.867}   &  0.531   &  0.554   &  0.584   &  \textbf{0.733}   &  0.684   &  0.639   &  0.725   &  \textbf{0.818} \\
    \midrule
    Overall & 0.930   &  0.887   &  \textbf{0.941}   &  0.929   &  0.856   &  0.816   &  0.885   &  \textbf{0.922}   &  0.719   &  0.672   &  0.722   &  \textbf{0.785}   &  0.835   &  0.792   &  0.849   &  \textbf{0.879}  \\
    \bottomrule
  \end{tabular}
  \label{tab:table_embel_settings}
\end{table*}

2. Impacts of the embedding strategies for incorporating special padding token(specifically `@`) within the text context on the visual-language model. The four embedding strategies evaluated are: no embedding, prefix embedding, suffix embedding, and mixed embedding (embedding `@` within the text).

The results, as summarized in Table \ref{tab:table_embel_settings}, indicate significant variability in the performance of the visual-language model based on the embedding method used for the special character `@`. The evaluation encompasses three main tasks: classification (CLS), captioning (CAP), and visual question answering (VQA), each exhibiting distinct trends in success rates across different targets.

When considering overall performance, hybrid embedding emerges as the most sustainable and effective strategy, achieving the highest average success rate across tasks. This method's flexibility in integrating the special character @ within the text appears to enhance the model's interpretative capabilities, particularly in more context-dependent scenarios such as Visual Question Answering (VQA).

Suffix embedding, on the other hand, demonstrates superior performance in classification tasks. This suggests that placing the special character at the end of the text helps maintain contextual integrity, thereby improving the model's ability to correctly classify inputs. The success of suffix embedding in classification tasks implies that the placement of special characters can significantly impact the effectiveness of the attack, with the suffix position causing the least disruption to the model's processing flow.

Interestingly, when examining the text concealment rate during the attack, the method without embedding shows the highest effectiveness. This approach does not introduce additional symbols such as @, which can sometimes alert the model or the user to the presence of an attack, thus maintaining a more natural and undetectable text flow. The non-embedded method also exhibits a relatively high success rate in attacks compared to the baseline, indicating that simplicity and subtlety in embedding can sometimes be more advantageous than complex embedding strategies.

\subsection{Example of cross prompt task}
\label{sec:appendix_2}

Example dataset of transfer attack text prompts excerpted from CroPA\cite{luo2024image}, divided into three categories: image classification(CLS), image captioning(CAP), and visual question answering(VQA).

\subsubsection{Examples of CLS task}

\begin{itemize}
    \item \textit{"If this image were turned into a jigsaw puzzle, what would the box label say to describe the picture inside?"}
    \item\textit{"Classify the content of this image."}
    \item\textit{"If you were to label this image, what label would you give?"}
    \item\textit{"What category best describes this image?"}
    \item\textit{"Describe the central subject of this image in a single word."}
    \item\textit{"Provide a classification for the object depicted in this image."}
    \item\textit{"If this image were in a photo album, what would its label be?"}
    \item\textit{"Categorize the content of the image."}
    \item\textit{"If you were to sort this image into a category, which one would it be?"}
    \item\textit{"What keyword would you associate with this image?"}
\end{itemize}

\subsubsection{Examples of CAP task}

\begin{itemize}
    \item \textit{"Describe the content of this image."}
    \item \textit{"What's happening in this image?"}
    \item \textit{"Provide a brief caption for this image."}
    \item \textit{"Tell a story about this image in one sentence."}
    \item \textit{"If this image could speak, what would it say?"}
    \item \textit{"Summarize the scenario depicted in this image."}
    \item \textit{"What is the central theme or event shown in the picture?"}
    \item \textit{"Create a headline for this image."}
    \item \textit{"Explain the scene captured in this image."}
    \item \textit{"If this were a postcard, what message would it convey?"}
\end{itemize}

\subsubsection{Examples of VQA task}

\begin{itemize}
    \item \textit{"Any cutlery items visible in the image?"}
    \item \textit{"Can you find any musical instruments in this image?"}
    \item \textit{"Does the image appear to be a cartoon or comic strip?"}
    \item \textit{"How many animals are present in the image?"}
    \item \textit{"Is a chair noticeable in the image?"}
    \item \textit{"How many statues or monuments stand prominently in the scene?"}
    \item \textit{"How many different patterns or motifs are evident in clothing or objects?"}
    \item \textit{"What is the spacing between objects or subjects in the image?"}
    \item \textit{"Would you describe the image as bright or dark?"}
    \item \textit{"What type of textures can be felt if one could touch the image's content?"}
\end{itemize}




\end{document}